\documentclass[prd,aps,amssymb
,nofootinbib,superscriptaddress,floatfix,showpacs]{revtex4}

\usepackage{amsmath}
\usepackage{graphicx} 
\usepackage{latexsym}
\usepackage{amsfonts}
\usepackage{url,hyperref}
\usepackage{subfig} 
\usepackage{bbm} 
\usepackage{verbatim}
\usepackage{bm}
\usepackage{color}
\newcommand{\beq}{\begin{equation}}
\newcommand{\eeq}{\end{equation}}
\def\bea{\begin{eqnarray}}
\def\eea{\end{eqnarray}}

\begin{document}

\title{Entanglement of two qubits in a relativistic orbit}
\author{Jason~Doukas}
\email[Email: ]{jasonad@yukawa.kyoto-u.ac.jp}
\affiliation{Department of Mathematics and Statistics, The University of Melbourne, Parkville, Victoria 3010, Australia.}
\affiliation{Yukawa Institute for Theoretical Physics, Kyoto University, Kyoto, 606-8502, Japan.}
\author{Benedict~Carson}
\affiliation{School of Physics, The University of Melbourne, Victoria 3010, Australia}

\begin{abstract}
The creation and destruction of entanglement between a pair of interacting two-level detectors accelerating about diametrically opposite points of a circular path is investigated. It is found that any non-zero acceleration has the effect of suppressing the vacuum entanglement and enhancing the acceleration radiation thereby reducing the entangling capacity of the detectors. Given that for large accelerations the acceleration radiation is the dominant effect, we investigate the evolution of a two detector system initially prepared in a Bell state using a perturbative mater equation and treating the vacuum fluctuations as an unobserved environment. A general function for the concurrence is obtained for stationary and symmetric worldlines in flatspace. The entanglement sudden death time is computed.  
\end{abstract}

\preprint{YITP-10-16}
\date{\today}
\maketitle

\section{Introduction}
\par It has been shown \cite{Summers1985,Summers1987} in algebraic quantum field theory that the vacuum state of a relativistic field is highly entangled. As by definition there are no real particles in a vacuum, entanglement defined here refers to the entangling effect two spacelike separated detectors would experience when placed into the vacuum. This inherently non-local effect is exhibited by the Feynman propagator which is non-vanishing between spacelike regions; a phenomena attributed to the exchange of virtual particles \cite{Reznik2005, Silman2005,Pachos2003}.

\par With an understanding of vacuum fields in inertial frames it is possible to progress to entanglement with the vacuum as observed by highly relativistic and non-inertial observers. Many such studies have now been performed  \cite{Gingrich2002, Alsing2003,Alsing2003a, Alsing2006, NicolasC.Menicucci2009,Doukas:2008ju,Fuentes-Schuller2005,Ball:2005xa,SergeMassar2006,Landulfo2009,Wang2009}. 

\par
\begin{figure}[h]
\centering
\includegraphics[scale=0.5]{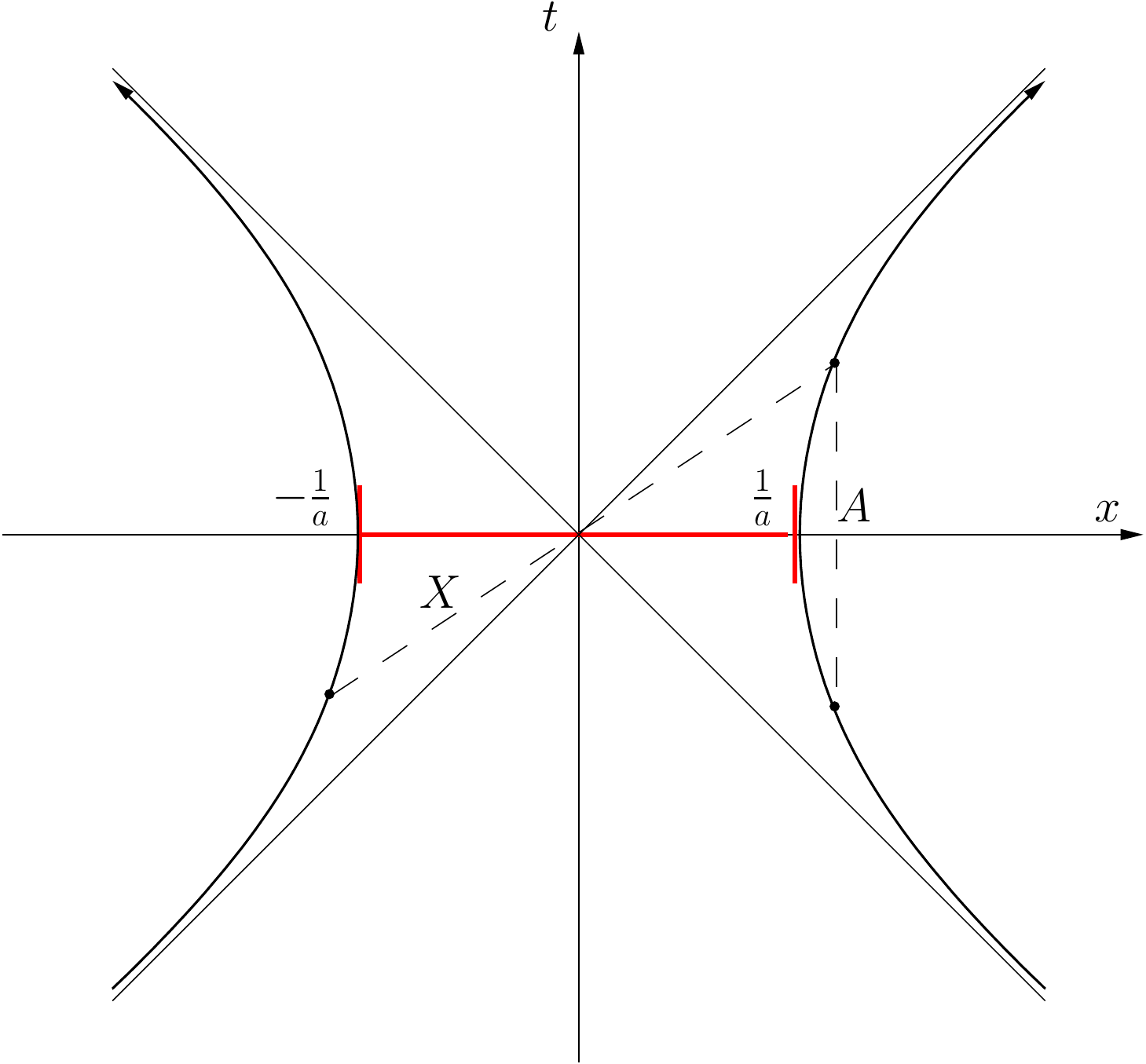}
\caption{Interactions with a massless scalar field of particle detectors collinearly accelerating in opposite directions. The amplitude $X$ depends on field correlations that exist between the spactime events of the two detectors while the amplitude $A$ depends on the field correlations that exist between events that occur at different times along the worldline of a single detector. At the origin the detectors are separated by a distance $\frac{2}{a}$.}
\label{fig:linear_acc}
\end{figure}

 In the case of linear acceleration one would expect the entanglement to be reduced by the interactions of the detectors with the Unruh bath \cite{Unruh1976}. However for the accelerated two-level particle detectors shown in figure \ref{fig:linear_acc}, it was found \cite{Reznik2003} that 
\begin{equation}
  |X|=e^{\frac{\pi \Omega}{2 a}} A
\end{equation}
where $a$ is the acceleration of the detectors and $\Omega$ the energy of the level separation.  The requirement that an initially unentangled state become entangled can be expressed as $|X|>A$. Thus, in the setup of figure \ref{fig:linear_acc} the detectors are always entangled.

\par  One can interpret $A$ as the transition probability for the two-level detector to flip from ground to excited state, it does not contain any interaction terms between the detectors and therefore does not provide any entanglement. As the acceleration is increased,  the transition probability per unit time approaches infinity. This is not surprising since the level flipping is enhanced by the Unruh radiation. What is surprising is that the amplitude $|X|$ (per unit time) also diverges. In other words, as the acceleration is increased, the cross detector vacuum interactions are enhanced. Since in this setup the detectors are separated by twice the inverse of the acceleration it is not clear whether it is the acceleration {\it per se} that increases $|X|$ or if the increase is merely an artifact of the detectors being closer together. We are thus motivated to explore other systems of non-inertial motion to clarify this issue. 
\par The obvious place to start would be to enlarge the number of independent parameters in the linearly accelerating case, so that the distance between the detectors at the origin could be varied independently of the acceleration. However, this is not the simplest case; one finds that the correlation function between detectors depends on both $\tau-\tau'$ and $\tau+\tau'$ in a non-factorizable way.  Ideally we would like to find non-inertial motion such that the single detector correlations are stationary \cite{Letaw1981,Obadia2007} and the cross detector correlations can be factorized into functions of $\tau-\tau'$ and $\tau+\tau'$. 

\par When the non-inertial motion is not strictly linear, the corresponding ``Unruh'' radiation seen by the detectors is known as acceleration radiation \cite{Hacian1985}. Typically on a stationary worldline the spectrum of vacuum fluctuations is not exactly Plankian but dependent on the motion of the detector \cite{Letaw1980,Kim1987}.

\par Regarding the physics, obvious experimental difficulties prevent the experimental interrogation of the vacuum in the linear case. Furthermore, the constant linear acceleration dilates the typical timescales involved to astronomically long magnitudes. However, for a particle following a circular path the acceleration
\begin{equation}
a=\frac{\gamma^2c^2}{R}
\end{equation}
is increased by a square of the Lorentz factor leading to a dramatic enhancement of the acceleration radiation effect, what is more since the speed is constant the time-dilation can be kept moderate; these points were first discussed in the context of measuring the Unruh temperature using single circulating electrons in \cite{Bell1983}, a recent elaboration on the depolarisation of electrons in storage rings and it's connection to acceleration radiation can be found in \cite{Unruh:1998gq, Leinaas1998}. There is also the possibility of experimentation with electrons brought to high circular accelerations in a Penning trap \cite{Rogers1988}. 

\par
\begin{figure}[h]
\centering
\includegraphics[scale=.7]{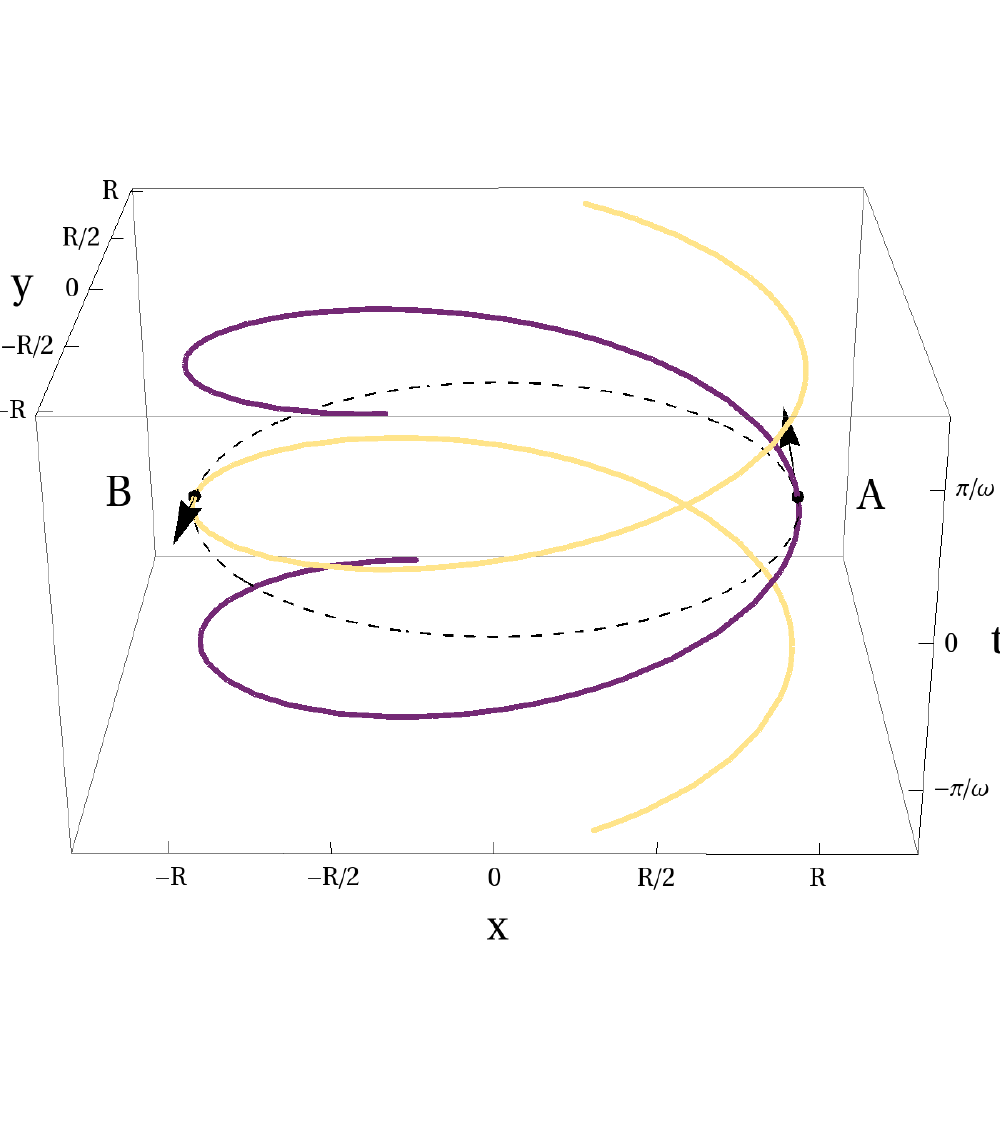}
\caption{Double helical structure of the particle detector worldlines, with radius $R$ and angular frequency $\omega$.}
\label{fig:circ_acc}
\end{figure}

 For the reasons mentioned this paper will investigate the quantum information between two two-level particle detectors rotating diametrically opposite along a circular path with angular frequency $\omega$, see figure \ref{fig:circ_acc}. The symmetry of the detectors' motion is found to greatly simplify the problem. Importantly, both correlation functions are stationary. Furthermore, since the detectors have equal speeds the energy splittings and interaction timescales are not Lorentz shifted relative to each other, as would be the case if one of the detectors were kept at rest. The detectors at any given time are a fixed distance apart and have a uniform acceleration pointing toward each other. This provides an ideal setup in which to investigate the effect of acceleration on vacuum entanglement. 

\par Section \ref{sec:averaged} will investigate the significance of the vacuum entanglement by adiabatically switching on the interaction using a gaussian window function. Section \ref{sec:dynamics} will focus on the density dynamics when the $A$ interactions dominate starting with a maximally entangled initial state and investigating it's evolution and the time dependence of it's concurrence in the inertial frame defined by the center of motion of the particles. Throughout this paper the notation used is $\hbar=c=1$, $\sigma_j$ where $j\in {x,y,z}$ are the Pauli matrices and $\sigma_0$ is the identity matrix, and the metric is $\text{diag}(+,-,-,-)$.  


\section{Entanglement with Window Functions} \label{sec:averaged}
\par The Hilbert space of the two circularly orbiting qubit particle detectors, $A$ and $B$, and a massless scalar field, 
$\phi$, is written as $\mathcal{H}=\mathcal{C}^2_A\otimes\mathcal{C}^2_B\otimes\mathcal{H}_{\phi}$. The Hamiltonian of the 
detectors, scalar field, and their interaction is given by $H=H_0+H_{\phi}+V$, where,
\begin{align}\label{eqn:hamiltonian}
 &H_0 =  -\frac{1}{2}\Omega(\sigma_z\otimes\mathbbm{1}+\mathbbm{1}\otimes\sigma_z)\notag\\
 &V = \eta(\tau)(\sigma_x\otimes\mathbbm{1}+\mathbbm{1}\otimes\sigma_x)\phi(x(\tau))
\end{align}
in which $H_{\phi}$ is the free Hamiltonian for the field $\phi$, and $\Omega$ is the energy difference between the two states of a qubit particle detector \cite{Audretsch1994}. For convenience the interaction is turned on at time $\tau=0$ using the Gaussian, $\eta(\tau)=\eta_0 e^{-\frac{\tau^2}{2\xi^2}}$ window function \cite{NicolasC.Menicucci2009,Sriramkumar:1994pb}, where $\xi$ is a fixed parameter used to control the interaction time and $\eta_0$ is the coupling. So far all of the above have been defined in the instantaneous rest frame (IRF) of the detectors, but it is more convenient to work in the inertial frame at the center of rotation and so we define\footnote{In fact $a'$ defined here is not the coordinate acceleration in the inertial frame which would be given by $\frac{dv}{dt}=R\omega^2=\gamma^2 a$. Instead $a'$ should be thought of as the change in coordinate velocity with respect to the change in proper time of a clock comoving with the detector, i.e., $a'=\frac{dv}{d\tau}$ .},
\begin{equation}
  \Omega'= \frac{\Omega}{\gamma}, \quad \eta'=\frac{\eta}{\gamma}, \quad \xi'=\gamma\xi, \quad a'=\frac{a}{\gamma}
\end{equation}
and the worldlines for the two detectors are,
\begin{eqnarray}
x_A&=&(t, R \cos \omega t, R \sin \omega t,0)\\
x_B&=&(t, -R \cos \omega t, -R \sin \omega t,0)
\end{eqnarray}
where,
\begin{equation}\label{eqn:omegaR}
  R=\frac{\gamma^2\beta^2}{a} \hspace{0.5cm}\text{and} \hspace{0.5cm} \omega=\frac{a}{\gamma^2\beta}.
\end{equation}
The positive frequency scalar field correlation functions,
\begin{equation}
 D^+(x_i(t),x_j(t'))=-\frac{1}{4\pi^2}\frac{1}{(x_i(t)-x_j(t'))^2},
\end{equation}
for the paths considered are,
\begin{align}\label{eqn:wightman}
 D^+(x_A(t),x_B(t'))=-\frac{1}{4\pi^2}\frac{1}{(t-t')^2-(2R)^2\cos^2(\omega(t-t')/2)},\notag\\
 D^+(x_A(t),x_A(t'))=-\frac{1}{4\pi^2}\frac{1}{(t-t')^2-(2R)^2\sin^2(\omega(t-t')/2)},
\end{align}
and equivalent expressions for the correlation functions with $A\leftrightarrow B$ interchanged.

\par Second order perturbation theory is used to calculate the final out density given it's initial state, $\rho(-\infty)=|\downarrow\downarrow\rangle\langle\downarrow\downarrow|$. One begins with the Schr\"{o}dinger equation for the density operator, $\dot{\rho}(t)=-i[V(t), \rho(t)]$. Integrating over $\dot{\rho}(t)$, 
and substituting the result back into the right hand side, then integrating a second time gives,
\begin{eqnarray}\label{eqn:rhointegral}
\rho(\infty)=\rho(-\infty)-i\int^{\infty}_{-\infty}[V(t'),\rho(-\infty)]dt'
-\int^{\infty}_{-\infty}dt'\int^{t'}_{-\infty}[V(t'),[V(t''),\rho(-\infty)]]dt''.
\end{eqnarray}
Assuming that the detectors and field are initially uncorrelated so that $\rho(-\infty)=\rho_s(-\infty)\otimes\rho_{\phi}(-\infty)$ 
and $\rho_{\phi}(-\infty)=|0\rangle\langle 0|$, and solving (\ref{eqn:rhointegral}) leads to the final out density,
\begin{eqnarray}\label{eqn:rhofinal}
\rho_s(\infty)=
\left(\begin{array}{cccc}
0 & 0 & 0 & X\\
0 & A & Y & 0\\
0 &Y^*& A & 0\\
X^*&0 & 0 & 1-2A 
\end{array}
\right)
\end{eqnarray}
where
\begin{eqnarray}
A&=&\int_{-\infty}^{\infty}dt'\int_{-\infty}^{\infty}dt''\eta'(t')\eta'(t'')e^{-i\Omega' (t'-t'')} D^+(x_A(t'),x_A(t''))\label{eqn:A}\\
X&=&-2\int^{\infty}_{-\infty} dt'\int^{t'}_{-\infty} dt''\eta'(t')\eta'(t'')e^{i\Omega' (t'+t'')}D^+(x_A(t'),x_B(t''))\label{eqn:X}\\
Y&=&2\text{Re}\int^{\infty}_{-\infty} dt'\int^{t'}_{-\infty} dt''\eta'(t')\eta'(t'')e^{-i\Omega' (t'-t'')}D^+(x_A(t'),x_B(t'')).
\end{eqnarray}

\par A necessary and sufficient condition for inseparability of (\ref{eqn:rhofinal}), and also for entanglement, is that 
the negativity (defined as the absolute value of the sum of the negative eigenvalues of the partial transposed matrix) be 
greater than zero \cite{PhysRevLett.77.1413,Horodecki1995340,PhysRevA.65.032314}. As discussed in the introductory example, this amounts to $|X|>A$.

\par
\begin{figure}[h]
\centering
\includegraphics[scale=0.8]{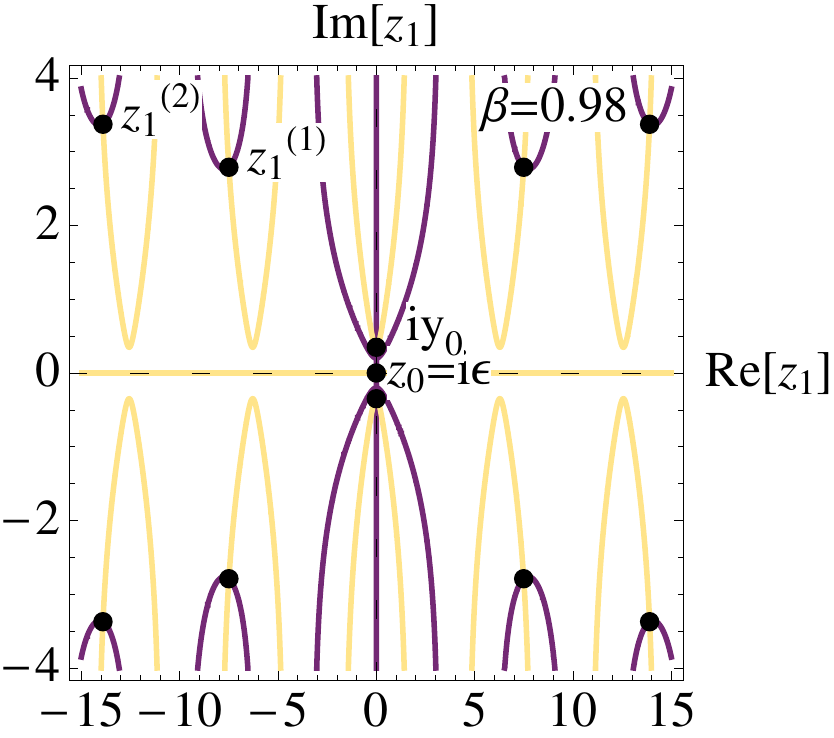}\hspace{1cm}
\includegraphics[scale=0.8]{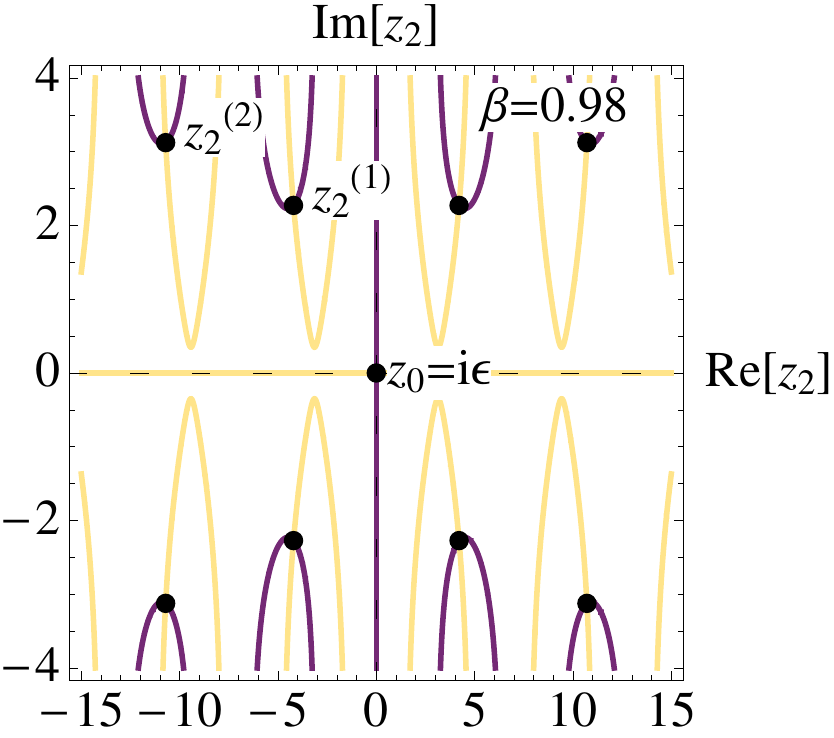}
\caption{Due to the symmetry of equations (\ref{eqn:Apoles}) the poles are symmetric about the real and imaginary axes. We define the independent poles to have positive imaginary part and negative real part, and label them in order of their distance away from the origin. All poles are first order except for $z_0$ which is a second order pole.}
\label{fig:APoles}
\end{figure}

 Equations (\ref{eqn:A}) and (\ref{eqn:X})  can now be solved for the circular correlation functions (\ref{eqn:wightman}).
The amplitude $A$ can be solved using contour integration, one finds poles distributed in the complex plane satisfying the equations
\begin{eqnarray}\label{eqn:Apoles}
  z_{1}&=&\beta \sin z_{1},\\ 
  z_{2}&=&-\beta \sin z_{2}. 
\end{eqnarray}
These equations are numerically solved in figure \ref{fig:APoles}. An approximate solution for $y_0$ can be obtained\footnote{Comparison with the exact numerical solution shows that this approximation is only valid in the relativistic regime, we found a useful ansatz in the $\beta\rightarrow0 $ limit was $y_0=\alpha_1(\beta) \text{arccosh}(\beta^{-1})$ where $1<\alpha(\beta)< \sqrt{3}$.}
\begin{equation}
  y_0\approx \sqrt{6(\beta^{-1}-1)}
\end{equation}
Defining the dimensionless variables
\begin{eqnarray}
  r=\frac{R}{\xi} \quad y=\Omega\xi \quad \alpha=a\xi.
\end{eqnarray}
and performing the contour integration we find
\begin{eqnarray}\label{eqn:Aresult}
  A&=&\frac{\eta_0^2}{4 \pi}\left\{ e^{-y^2}-\sqrt{\pi} y \text{erfc}(y)+\frac{\sqrt{\pi}} {4 \gamma^2}\sqrt{\frac{\alpha}{r}}\frac{e^{\frac{y_0^2r}{\alpha}}}{y_0(y_0\coth y_0-1)}\left(e^{2 y_0 y \sqrt{\frac{r}{\alpha}}}\text{erfc}\left(y_0\sqrt{\frac{r}{\alpha}}+y\right)+e^{-2y_0 y\sqrt{\frac{r}{\alpha}}}\text{erfc}\left(y_0\sqrt{\frac{r}{\alpha}}-y\right)\right)\right.\nonumber\\ 
  &+& \left.\frac{\sqrt{\pi}}{2\gamma^2}\sqrt{\frac{\alpha}{r}} \text{Im} \sum_{z_p\in\cup\{z_1^{(k)},z_2^{(k)}\}}\frac{e^{-\frac{z_p^2 r}{\alpha}}}{z_p(1-z_p\cot z_p)}\left(e^{-2iz_p y\sqrt{\frac{r}{\alpha}}}\text{erfc}(-iz_p\sqrt{\frac{r}{\alpha }}+y)+e^{2iz_p y\sqrt{\frac{r}{\alpha }}}\text{erfc}(-iz_p\sqrt{\frac{r}{\alpha}}-y) \right) \right\}
\end{eqnarray}
where the gamma factor  
\begin{eqnarray}
\gamma^2&=&r\alpha+1
\end{eqnarray}
has been kept explicit for presentational purposes. Also the poles are dependent on $\beta$ which is
\begin{eqnarray}
\beta&=&(1+\frac{1}{r\alpha})^{-1/2}
\end{eqnarray}
in terms of the dimensionless parameters.
\par
\begin{figure}[h]
\centering
\includegraphics[scale=.8]{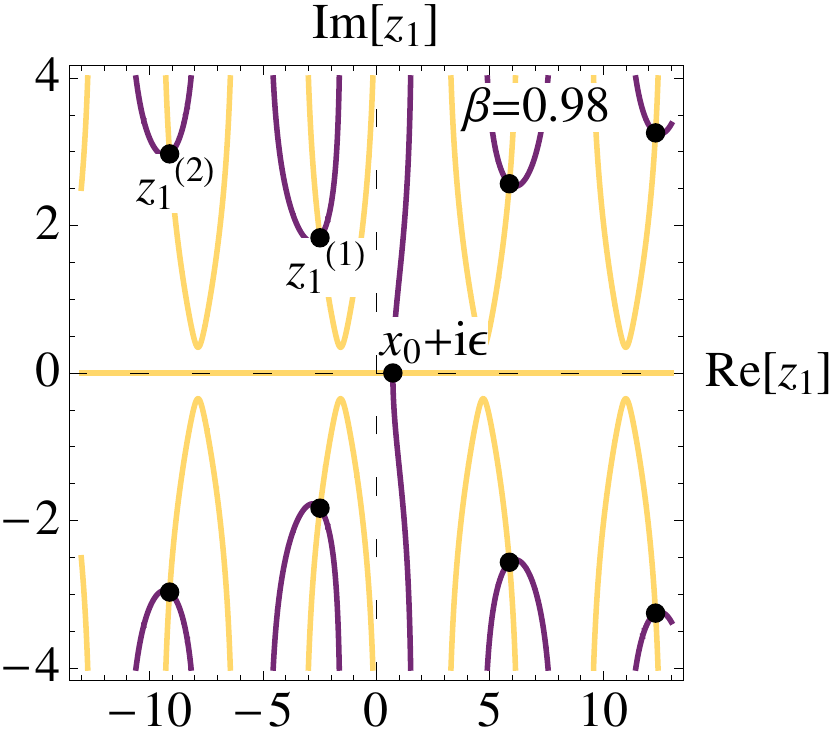}\hspace{1cm}
\includegraphics[scale=.8]{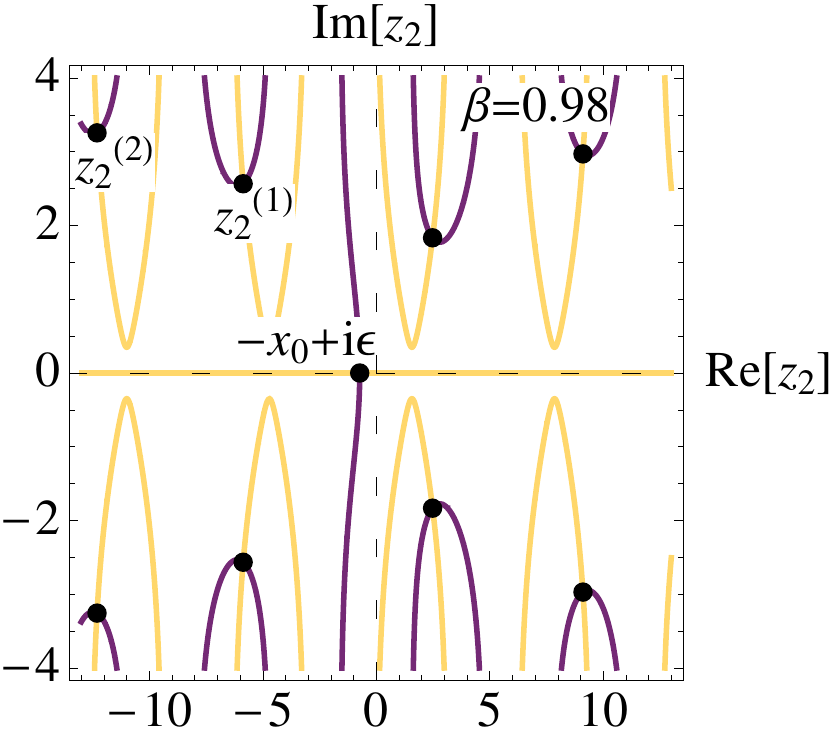}
\caption{Due to the symmetry of equations (\ref{eqn:xpoles}) the poles can be divided into sets of four which are related to each other by complex conjugation and negation. We define the independent poles to have positive imaginary part and negative real part, and label them in order of their distance away from the origin. }
\label{fig:XPoles}
\end{figure}

 The real part of $X$ can also be solved using contour integration, one finds poles distributed in the complex plane satisfying the equations
\begin{eqnarray}\label{eqn:xpoles}
  z_{1}&=&\beta \cos z_{1},\\ 
  z_{2}&=&-\beta \cos z_{2}. 
\end{eqnarray}
These equations are numerically solved in figure \ref{fig:XPoles}. An approximate solution for $x_0$ can be obtained
\begin{equation}
  x_0\approx -\beta^{-1}+\sqrt{\beta^{-2}+2},
\end{equation}
but the remaining poles must be determined numerically. The imaginary part of X can be computed directly (without the use of contour integration) adding both parts together we find that
\begin{eqnarray}\label{eqn:Xresult}
  X&=&-\frac{1}{4\sqrt{\pi}}\frac{\eta_0^2}{\gamma^2}\sqrt{\frac{\alpha}{r}}e^{-y^2}  \left(\frac{e^{-\frac{x_0^2r}{\alpha}}(\text{erfi}(x_0\sqrt{\frac{r}{\alpha}})-i)}{2x_0(1+x_0\tan x_0)} + \text{Im}\sum_{z_p\in\cup\{z_1^{(k)},z_2^{(k)}\}}\frac{e^{-\frac{z_p^2r}{\alpha}}\left(1+i \text{erfi}(z_p\sqrt{\frac{r}{\alpha}})\right)}{z_p(1+z_p\tan z_p)} \right).
\end{eqnarray}

\par By increasing the size of the window function, we can obtain the transition probability per unit time found in \cite{Bell1983} \footnote{The choice of limit on the left hand side can be understood in the following way: the authors in \cite{Bell1983} define the transition rate per unit time by dividing the transition probability by $\int_{\infty}^{\infty}d\frac{1}{2}(t+t')$ which corresponds in the case with a window function to dividing by a factor of $\lim_{\xi\rightarrow\infty}\frac{1}{2}\int_{-\infty}^{+\infty}e^{-\frac{y^2}{4\xi^2}}dy=\sqrt{\pi}\xi$.} 
\begin{equation}\label{eqn:Alongtime}
  \lim_{\xi\rightarrow\infty}\frac{A}{\sqrt{\pi}\xi}=\frac{\eta_0^2a}{8\sqrt{3}\pi}e^{-\frac{\sqrt{12}\Omega}{a}}.
\end{equation}
On the other hand we find that
\begin{equation}\label{eqn:Xlongtime}
  \lim_{\xi\rightarrow\infty}X=0,
\end{equation}
this can be understood by observing that the rate per unit time of $X$ oscillates about zero and in the long time limit averages to zero.    
\par It is also satisfying that by replacing, $\alpha=\gamma^2\beta^2/r$, and taking the the slow motion limit our equations reduce to the inertial case \cite{NicolasC.Menicucci2009}\footnote{Upon making the correspondence in notation $2 R\rightarrow L$ and $\xi\rightarrow\sigma$. There is also the appearance of the imaginary term in $X$ not present in \cite{NicolasC.Menicucci2009}, we have confirmed via private communication with the authors that the expression as we have it here is correct.}:
\begin{eqnarray}
  \lim_{\beta\rightarrow 0} A &=& \frac{\eta_0^2}{4 \pi} \left(e^{-y^2}-\sqrt{\pi} y \text{erfc} (y)\right),\\
\lim_{\beta\rightarrow 0}X &=& -\frac{\eta_0^2}{4\sqrt{\pi}}\frac{1}{2 r}e^{-r^2-y^2}
	\left(\text{erfi}(r)-i\right).
\end{eqnarray}
We can now investigate the large acceleration limit while keeping the radius fixed, we find
\begin{eqnarray}
  \lim_{\alpha\rightarrow\infty}A&=& \frac{\eta_0^2\alpha}{8 \sqrt{3\pi}}\label{eqn:Alargea}\\
  \lim_{\alpha\rightarrow\infty}|X|&=&0.42 \frac{\eta_0^2}{4\sqrt{\pi}}\frac{e^{-y^2}}{r^{3/2}} \frac{1}{\sqrt{\alpha}}\label{eqn:Xlargea}
\end{eqnarray}
where the decimal number in $|X|$ results from summing the contributions of the $z_1^{(k)}$ and $z_2^{(k)}$ poles up to $k=10$ .
\par
\begin{figure}[h]
\centering
\includegraphics[scale=.7]{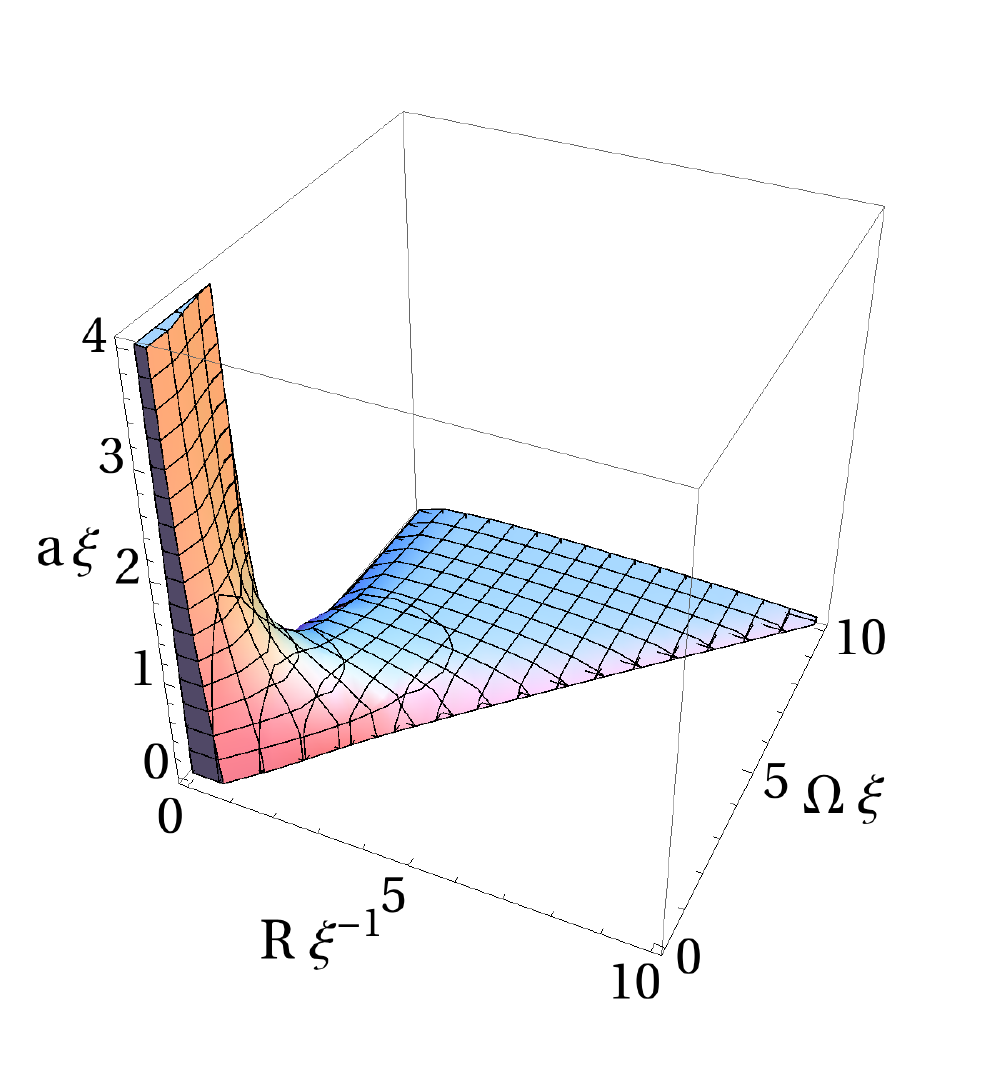}
\caption{Plot of the entanglement volume defined by the condition $|X|>A$. This plot includes the pole contributions to equations (\ref{eqn:Xresult}) and (\ref{eqn:Aresult})  up to $k=10$. As the acceleration increases the entanglement volume is squeezed toward the $\alpha-y$ plane, thus at large acceleration the detectors only entangle for very small radii. In the opposite limit ($a\xi\rightarrow 0$) the general features of the inertial case \cite{NicolasC.Menicucci2009} can be seen; a line in the $r-y$ plane separates entangled from non-entangled regions.}
\label{fig:rotneg}
\end{figure}

 Thus at fixed radius the cross detector correlations are reduced like the inverse square root of the acceleration. Furthermore, the $A$ amplitude diverges proportionally with $\alpha$. Both of these processes combined make it very difficult for any entanglement to be gained by detectors with large acceleration. 
The entanglement region is plotted in figure \ref{fig:rotneg}.
\par Having shown that circular acceleration suppresses the production of vacuum entanglement, in the next section we will investigate the loss of entanglement due to the acceleration radiation.  We calculate the entanglement lifetime of a pair of detectors that are initially entangled in a Bell state assuming for simplicity that the cross detector interactions responsible for the creation of vacuum entanglement can be ignored.


\section{Entanglement dynamics}\label{sec:dynamics}
\par In this section the dynamics of the two entangled detectors in the presence of a real scalar field are investigated. The method follows that given in \cite{Doukas:2008ju}. Initially, the density operator of the two detectors are prepared in a maximally entangled state, called the Bell state $\rho_{Bell}=\frac{1}{4}(\mathbbm{1}\otimes\mathbbm{1}+\sigma_x\otimes\sigma_x-\sigma_y\otimes\sigma_y+\sigma_z\otimes\sigma_z)$, which corresponds to the pure state $(|\downdownarrows\rangle+|\upuparrows\rangle)/\sqrt{2}$ written as a density matrix. As mentioned in the last section, since the vacuum entanglement is suppressed by acceleration it will be safe to ignore cross detector interactions. Each detector can then be considered in isolation before adding the two together to solve for the total density operator. For ease of notation we suppress the factors of $\otimes \sigma_0$ so that for example $\sigma_+$ should read $\sigma_+\otimes\sigma_0$. Utilising the Hamiltonian in equation (\ref{eqn:hamiltonian}), the equation of motion to second order for detector $A$ in the Schr\"{o}dinger picture is
\begin{align}\label{eqn:rhodifferential}
  \dot{\rho}_s &=-i[H_0,\rho_s]-{\eta'_0}^{2}\Big\{\sum_i\int_0^{t}dt' e^{i\Omega'(t'-t)}
(\sigma_i\sigma_+\rho_s-\sigma_+\rho_s\sigma_i)D^+(x_A(t),x_A(t'))\notag\\
&\hspace{2cm}+\sum_i\int_0^{t}dt' e^{-i\Omega'(t'-t)}
(\sigma_i\sigma_-\rho_s-\sigma_-\rho_s\sigma_i)D^+(x_A(t),x_A(t'))+h.c.\Big\}.
\end{align}
The correlation functions result from tracing over the unmeasured vacuum field, where 
$Tr[\rho_{\phi}\phi\phi']$ reduces to $D^+(x_A(t),x_A(t'))$ and is the same as equation (\ref{eqn:wightman}).

\par In order to simplify the second order equation (\ref{eqn:rhodifferential}) it is convenient to define $f_1=(I_+ + I_-)/2$, 
$f_2=-i(I_+ - I_-)/2$, $d_1=1$, and $d_2=0$, in which
\begin{equation}\label{eqn:waveintegral}
 I_\pm = \int_0^{t}dt' e^{\pm i\Omega'(t'-t)}D^+(x_A(t),x_A(t')).
\end{equation}
As this integral is sharply peaked and drops of quickly outside the domain of integration the Markovian approximation can be 
used to extend the integral to $[0,\infty]$. By extending the domain of integration in this way the real part 
of the integral equation (\ref{eqn:waveintegral}) is even, and can be solved by closing a loop integral in the upper (lower) half 
of the complex plane for $I_+$ ($I_-$). The imaginary part is odd, diverging at the origin, and will need to be removed.

\par After some simplification, the new notation gives the equation,
\begin{equation}
  \dot{\rho}_s =-i[H_0,\rho_s]-{\eta'_0}^2\sum_{i,j}\frac{1}{2}\Big\{(f_j d_i +f_i^*d_j)\left(\{\sigma_i\sigma_j,\rho_s\}
-2\sigma_j\rho_s\sigma_i\right)+(f_j d_i -f_i^*d_j)[\sigma_i\sigma_j,\rho_s]\Big\}.
\end{equation}
The last term, $(f_j d_i -f_i^*d_j)[\sigma_i\sigma_j,\rho_s]$, when expanded is non-zero only for $j=2$ and becomes $-i\text{Im} (I_+-I_-)[\sigma_z,\rho_s]/2$, 
which contains the divergent part of the integral (\ref{eqn:waveintegral}). To remove the divergence the free Hamiltonian is 
redefined as $H'=H_0+\frac{ {\eta'_0}^2}{2} \text{Im}(I_+-I_-)\sigma_z$; the master equation now reading as,
\begin{equation}
  \dot{\rho}_s=-i[H',\rho_s]-\frac{ {\eta'_0}^2}{2}\sum_{i,j}(f_j d_i+f_i^*d_j)(\{\sigma_i\sigma_j,\rho_s\}-2\sigma_j\rho_s\sigma_i)
\end{equation}
where the energy separation between the states of the detector is renormalized as $\omega'$.

\par Further simplification can be achieved by switching to the interaction picture in the basis $\sigma_{\pm}$ to get,
\begin{align}
  \dot{\rho}_I=&{\eta'_0}^2 \text{Re}(I_-) (2\sigma_-\rho_I\sigma_+-\{\sigma_+\sigma_-,\rho_I\}) + 
	  \eta_0^2 \text{Re}(I_+) (2\sigma_+\rho_I\sigma_--\{\sigma_-\sigma_+,\rho_I\})\notag\\
&\hspace{2cm}+\eta_0^2 e^{2i\omega't}(I_++I_-^*)\sigma_+\rho_I\sigma_+
  +\eta_0^2 e^{-2i\omega't}(I_-+I_+^*)\sigma_-\rho_I\sigma_-.
\end{align}
The last two terms are rapidly oscillating due to the new renormalised energy, and can be dropped in the rotating wave approximation \cite{Agarwal1973}. Following the same procedure for detector $B$ and adding the two together results in the Lindblad form \cite{Banks1984} of the total master equation,
\begin{equation}
 \frac{d\rho}{dt}=\sum_j(2L_j\rho L_j^{\dagger}-\{L_j^{\dagger}L_j,\rho\})
\end{equation}
where the operators are,
\begin{align}
 L_1 &= \eta_0 \sqrt{\text{Re} (I_-)}(\sigma_-\otimes\mathbbm{1})&
 L_2 &= \eta_0 \sqrt{\text{Re} (I_+)}(\sigma_+\otimes\mathbbm{1})\notag\\
 L_3 &= \eta_0 \sqrt{\text{Re} (I_-)}(\mathbbm{1}\otimes\sigma_-)&
 L_4 &= \eta_0 \sqrt{\text{Re} (I_+)}(\mathbbm{1}\otimes\sigma_+).
\end{align}
In order to solve the above system of differential equations the density operator is expanded via $\rho=\sum_{i,j}r_{ij}\sigma_i\otimes\sigma_j$, 
and substituted back into the Lindblad equation. As the coefficient, $r_{ij}$, is symmetric when interchanging indices it is 
sufficient to look only at the top diagonal half of elements, which are,
\begin{align}
 \dot{r}_{00}&=0&
 \dot{r}_{zz}&=2{\eta'_0}^2(\text{Re}(I_- - I_+) r_{0z}-2\text{Re}(I_- + I_+)r_{zz}+\text{Re}(I_- - I_+) r_{z0})\notag\\
 \dot{r}_{0x}&=-{\eta'_0}^2 r_{0x}\text{Re}(I_+ + I_-)&
 \dot{r}_{0z}&=2{\eta'_0}^2 (r_{00}\text{Re}(I_- -I_+)-r_{0z}\text{Re}(I_+ + I_-))\notag\\
 \dot{r}_{0y}&=-{\eta'_0}^2 r_{0y}\text{Re}(I_+ + I_-)&
 \dot{r}_{xy}&=-2{\eta'_0}^2\text{Re}(I_- +I_+)r_{xy}\notag\\
 \dot{r}_{xx}&=-2{\eta'_0}^2\text{Re}(I_+ + I_-)r_{xx}&
 \dot{r}_{xz}&={\eta'_0}^2(2\text{Re}(I_- -I_+)r_{x0}-3\text{Re}(I_+ + I_-)r_{xz})\notag\\
 \dot{r}_{yy}&=-2{\eta'_0}^2 r_{yy}\text{Re}(I_+ + I_-)&
 \dot{r}_{yz}&={\eta'_0}^2(2\text{Re}(I_- -I_+)r_{y0}-3r_{yz}\text{Re}(I_+ + I_-)).
\end{align}
Solving for $r$ with the density operator initially being in the Bell state, the only non-zero elements are,
\begin{align}\label{eqn:rtimedependence}
 r_{00} &= \frac{1}{4}&
 r_{xx} &= -r_{yy} = \frac{1}{4}e^{-t/T_2}\notag\\
 r_{0z} &= r_{z0} = \frac{1}{4}\delta(1-e^{-t/T_2})&
 r_{zz} &= \frac{1}{4}\delta^2(1-2e^{-t/T_2}+(1+1/\delta^2)e^{-t/T_1}),
\end{align}
where 
\begin{eqnarray} \label{eqn:timescaledefs}
  T_1^{-1}&=&4 {\eta'_0}^2\ \text{Re}(I_- + I_+)\\
  T_2^{-1}&=&T_1^{-1}/2\label{eqn:relax2}
\end{eqnarray}
are the relaxation and dephasing times respectively. For convenience we have also defined 
\begin{equation}\label{eqn:deltadef}
  \delta=\frac{\text{Re}(I_- - I_+)}{\text{Re}(I_- + I_+)}.
\end{equation}
The integrals (\ref{eqn:waveintegral}) are solved by expanding the denominator in powers of $(t'-t)$ to give,
\begin{align}
 \text{Re}(I_-)&=\frac{\gamma^2}{4\pi}\left(\Omega' + \frac{a'}{4\sqrt{3}}
   e^{-\frac{2\sqrt{3}\ \Omega'}{a'}}\right)\notag\\
 \text{Re}(I_+)&=\text{Re}(I_-)-\frac{\gamma^2}{4\pi}\Omega'.
\end{align}

\begin{figure}[t]
\centering
\includegraphics[scale=0.7]{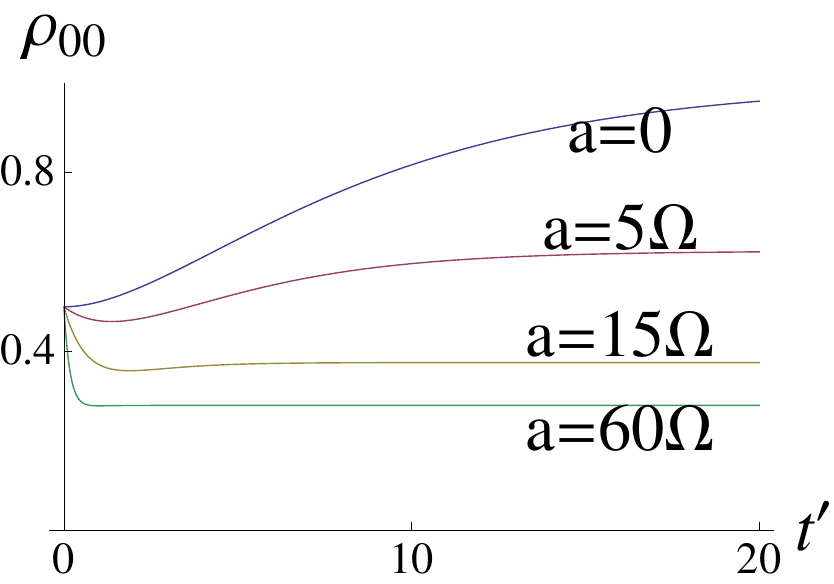}
\includegraphics[scale=0.7]{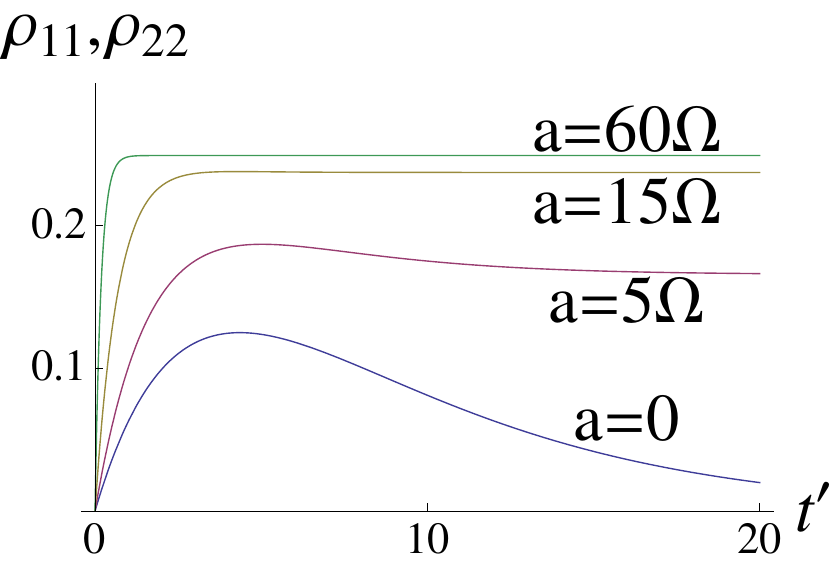}
\includegraphics[scale=0.7]{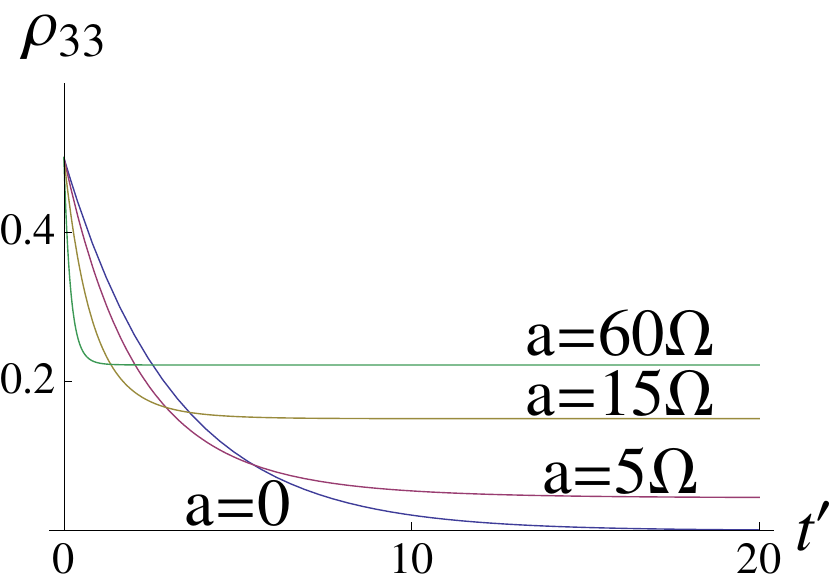}
\caption{The diagonal elements of the density operator for a given acceleration, $a$, are plotted as a function of rescaled time $t'=\frac{\eta_0^2 \Omega t}{\gamma}$.} 
\label{fig:sapprox}
\end{figure}

Substituting the results in (\ref{eqn:rtimedependence}) back into the expansion $\rho_s=\sum_{i,j}r_{ij}\sigma_i\otimes\sigma_j$ gives,
\begin{align}\label{eqn:finaldensity}
 \rho_{00}(t)&= \frac{1}{4}(e^{-t/T_1}+(1+\delta(1-e^{-t/T_2}))^2)\notag\\
 \rho_{11}(t)&= \rho_{22}(t)=\frac{1}{4}(1-e^{-t/T_1}-\delta^2(1-e^{-t/T_2})^2)\notag\\
 \rho_{33}(t)&= \frac{1}{4}(e^{-t/T_1}+(1-\delta(1-e^{-t/T_2}))^2)\notag\\
 \rho_{03}(t)&= \rho_{30}(t) = \frac{1}{2}e^{-t/T_2}
\end{align}
The $\rho_{00}(t)$ component is the fraction of an ensemble of detector system pairs which would have both detectors in the ground state, while $\rho_{33}(t)$ corresponds to both detectors being in the excited state. The remaining diagonal components, $\rho_{11}(t)$ and $\rho_{22}(t)$, are the fraction of detector pairs with one detector in the ground state and one in the excited state. These matrix elements are plotted in figure \ref{fig:sapprox} for various choices of acceleration.


\par Taking the limit as time goes to infinity, from equation (\ref{eqn:finaldensity}) the final out density reduces to,
\begin{eqnarray}\label{eqn:rhoequilibrium}
\rho_s(\infty)=
\left(\begin{array}{cc}
\frac{1}{2}(1+\delta) & 0 \\
0 & \frac{1}{2}(1-\delta)
\end{array}
\right)\otimes
\left(\begin{array}{cc}
\frac{1}{2}(1+\delta) & 0 \\
0 & \frac{1}{2}(1-\delta)
\end{array}
\right).
\end{eqnarray}
This can be expressed as the thermal state
\begin{equation}
  \rho_s(\infty)=\frac{1}{\text{Tr}{e^{-\beta_{\text{eff}}H_0}}}e^{-\beta_{\text{eff}}H_0}, 
\end{equation}
implying the detectors have reached thermal equilibrium with the scalar field at an effective temperature given by 
\begin{equation}
  T_{\text{eff}}=\frac{\Omega}{\ln \left( \frac{\rho_{\downarrow}}{\rho_{\uparrow}}\right)},
\end{equation}
where the population of the ground state to excited state of an individual detector is,
\begin{equation}
 \frac{\rho_{\downarrow}}{\rho_{\uparrow}}=\frac{1+\delta}{1-\delta}=1+ \frac{4\sqrt{3}\Omega}{a}e^{2\sqrt{3}\Omega /a},
\end{equation}
the same as for a single detector in circular motion \cite{Bell1983, Unruh:1998gq}. In terms of the effective temperature we can rewrite equations (\ref{eqn:timescaledefs}) and (\ref{eqn:deltadef}) as
\begin{eqnarray}
  T_1^{-1} &=& \frac{ \eta_0^2\Omega}{\pi\gamma\delta} \\
  \delta   &=& \tanh\frac{\beta_{\text{eff}}\Omega}{2}
\end{eqnarray}

\par
\begin{figure}[h]
\centering
\includegraphics{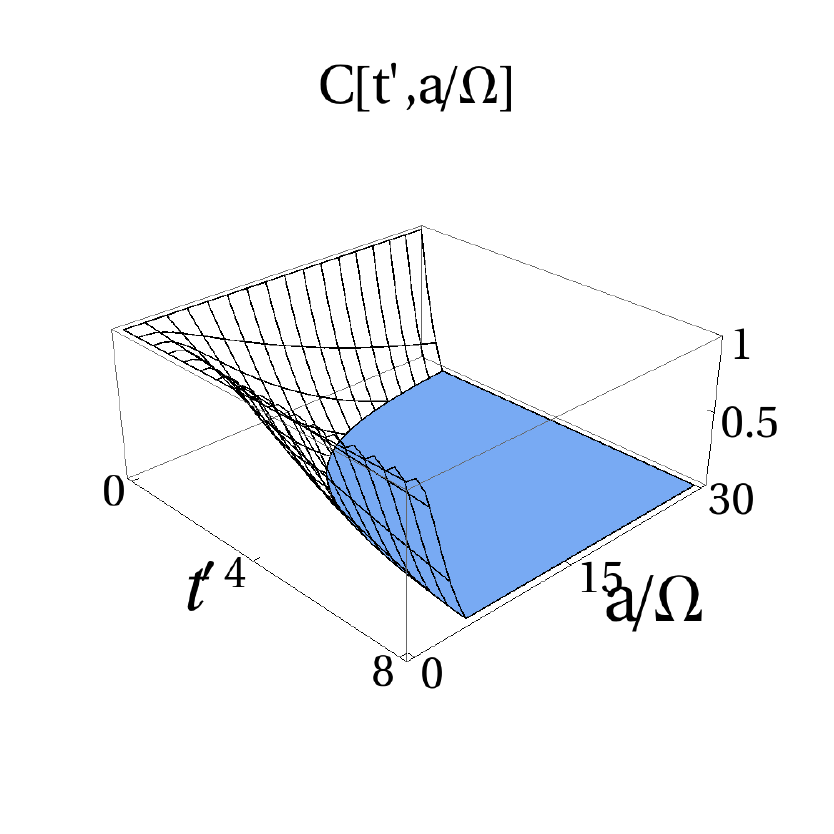}
\caption{The concurrence profile for the system of two detectors accelerating about helical worldlines, where again $t'=\eta_0^2\Omega/\gamma $ is the rescaled time and the acceleration is rescaled by the energy separation, $\Omega$.}
\label{fig:concurrence}
\end{figure}

 The density (\ref{eqn:rhoequilibrium}) in the long time limit is clearly separable and thus all the entanglement initially shared by the detectors is eventually lost. To quantitatively understand how the entanglement evolves with time the concurrence, \cite{Wootters1998}, is calculated to give a value between zero and one, corresponding to no entanglement and complete entanglement respectively. Concurrence is given by,
\begin{equation}
 C(\rho)=\max\{\lambda_1-\lambda_2-\lambda_3-\lambda_4,0\}
\end{equation}
where $\{\lambda_1,\lambda_2,\lambda_3.\lambda_4\}$ are the ordered positive square roots of the eigenvalues of the matrix 
$M=\rho\ (\sigma_y\otimes\sigma_y)\rho^*\ (\sigma_y\otimes\sigma_y)$. From (\ref{eqn:finaldensity}), the concurrence for the 
two detectors in circular motion is,
\begin{equation}\label{eqn:concurrence}
  C(\rho)=\max\{-(1-\delta^2)(\frac{1}{2}-e^{-t/T_2})+\frac{1}{2}(1+\delta^2)e^{-t/T_1},0\}
\end{equation}
and is plotted as a function of both time and acceleration in figure \ref{fig:concurrence}.

The curve along which the concurrence is zero is given by,
\begin{equation}\label{eqn:entlifetime}
  t_{\text{esd}}=T_2\ln\left(\frac{1+\delta^2}{\sqrt{2(1-\delta^2)}-1+\delta^2}\right).
\end{equation}
This behaviour is another example of what has been termed entanglement sudden death (ESD) \cite{Landulfo2009,Yu30January2009}. It can be seen that for zero acceleration ($\delta=1$) the concurrence decays exponentially with time, but for finite accelerations there is a finite time before which the concurrence completely disappears. 
\par As the acceleration increases and becomes much larger than the energy difference between states, the time taken for concurrence to go to zero is 
\begin{equation}
  t_{\text{esd}}\sim4\pi\sqrt{3}\ln\left(\frac{1}{\sqrt{2}-1}\right)\frac{\gamma}{\eta_0^2}\frac{1}{a},
\end{equation}
The $\gamma$ factor is due to the time dilation in transforming from the instantaneous rest frame to the inertial frame. As mentioned in the introduction, this dilation is less sever than that which occurs in the linear case, where the transformation is essentially exponential \cite{Doukas:2008ju}. 

\par Careful consideration of the steps involved in arriving at equations (\ref{eqn:concurrence}) and (\ref{eqn:entlifetime}) indicate that they will hold for any pair of two-level detector systems (assuming the cross interactions are negligible) moving on symmetric and stationary \cite{Letaw1981} worldlines.

\section{Conclusion}
\par This paper has considered two qubit detectors orbiting a common point in the presence of an unmeasured massless scalar field. This setup provides an ideal situation in which to study the vacuum entanglement for detectors undergoing uniform acceleration separated by some constant distance. 
\par We have found that acceleration suppresses entanglement between detectors by reducing the amplitude of the cross term like the inverse square root of the acceleration (\ref{eqn:Xlargea}) and increasing the level-flip transition amplitude linearly with the acceleration (\ref{eqn:Alargea}). Although our results pertain specifically to circular motion we conjecture that the way in which these amplitudes depend on the acceleration is universal to all non-inertial motions of detectors interacting with a free massless scalar field in flatspace. This conjecture can be justified in the following way: at any given time the detectors are a fixed distance apart and have fixed acceleration vectors pointing inward, we expect the direction of these acceleration vectors to be irrelevant to the local distortions of the field and therefore to the entangling behaviour of the field on the detectors.

\par  Having established that the vacuum interactions could be ignored for large accelerations we then studied the dynamics of the two detectors initialized in a Bell state. For large acceleration compared to the energy gap of the detector levels the ESD time (\ref{eqn:entlifetime}) was related to the relaxation time $T_2$ (\ref{eqn:relax2}) by
\begin{equation}
  t_{\text{esd}} \sim 0.881 T_2
\end{equation}
where $T_2$ was found to be inversely proportional to the acceleration. This indicates that it might be easier to observe the radiation effect by measuring the entanglement directly rather than the components of density matrix.  

\par We expect our expressions for the concurrence to be useful in other studies of non-inertial entanglement when the worldlines are symmetric and stationary. In particular, these will hold for the case of linearly accelerating worldlines with an independent distance parameter \footnote{In preparation.}.  


%
%
\acknowledgments
\par Financial support from the Australian Research Council via its support for the Centre of Excellence for Mathematics and Statistics of complex systems is gratefully acknowledged by JD. JD was supported by the Japan Society for the Promotion of Science (JSPS), under fellowship no. P09749.
\bibliography{rqi}{}
\end{document}